\documentclass[a4paper, 10pt, oneside, onecolumn, notitlepage, final]{article}
\usepackage[text={160mm, 240mm}, left=25mm, vmarginratio=1:1]{geometry}
\usepackage{graphicx}
\usepackage{amssymb, amsfonts, amsmath, amsthm, mathtools, mathrsfs}
\usepackage{upgreek}
\usepackage[usenames, dvipsnames]{xcolor}
\usepackage[colorlinks, linkcolor=blue, anchorcolor=red, citecolor=blue]{hyperref}
\usepackage[backend=biber, sorting=none, maxbibnames=3, minbibnames=3, style=numeric-comp]{biblatex}
\usepackage{braket} 
\usepackage{authblk} 
\usepackage{orcidlink}
\usepackage{abstract}
\usepackage{array}
\usepackage{bigstrut}
\usepackage{titlesec}
\usepackage[T1]{fontenc}
\usepackage[utf8]{inputenc}
\usepackage{lmodern}
\usepackage[english]{babel}

\titleformat{\section}{\large\bfseries\centering}{\thesection}{0.5 em}{}
\titleformat{\subsection}{\normalsize\bfseries\centering}{\thesubsection}{0.5 em}{}

\begin{document}

\title{\bf Responses for one-dimensional quantum spin systems via tensor networks}

\author{\normalsize Jiayin Gu\orcidlink{0000-0002-9868-8186}\thanks{\texttt{gujiayin@njnu.edu.cn}}}
\affil{\normalsize School of Physics and Technology, Nanjing Normal University, Nanjing 210023, China}

\date{}
\maketitle

\begin{abstract}
Tensor networks are adopted to calculate the responses for one-dimensional quantum spin systems that are initially in thermal equilibrium. The Ising chain in mixed transverse and longitudinal fields is used as the benchmarking system. The linear and second-order responses of the magnetization in $z$-direction induced by the time-dependent force conjugated with the magnetization in $x$-direction are calculated. In addition, the magnetization in $z$-direction is also exactly calculated in response to this excitation. As expected, the first two responses are shown to be excellent corrections to the equilibrium magnetization in $z$-direction when the excitation is weak. This result represents an illustrative example of the response theory for nontrivial quantum many-body systems.
\end{abstract}

\section{Introduction}

\par Tensor networks have been proven especially capable of dealing with quantum many-body systems~\cite{Montangero_2018, Orus_NatRevPhys_2019}. They were initially used to study the ground state properties of one-dimensional quantum system with density matrix renormalization group (DMRG)~\cite{White_PhysRevLett_1992, White_PhysRevB_1993} based on matrix product states (MPS)~\cite{Schollwock_AnnPhys_2011}, and subsequently extended to study the dynamics of quantum many-body systems out of equilibrium\cite{Eisert_NatPhys_2015, Paeckel_AnnPhys_2019}. In recent years, tensor networks find further applications in the thriving field of quantum thermodynamics~\cite{Deffner_2019, Strasberg_2021}. For example, they have been used to simulate the strongly interacting quantum thermal machines~\cite{Brenes_PhysRevX_2020}, to study the heat transfer in non-Markovian open quantum systems~\cite{Popovic_PRXQuantum_2021}, to calculate the work statistics for quantum spin chains~\cite{Gu_PhysRevRes_2022, Lin_PhysRevRes_2024}.

\par However, the birth of quantum thermodynamics dates back to the fifties in the last century when the response theory was established. This theory deals with the responses of a system to additional excitations in the Hamiltonian, and connects nonequilibrium properties with equilibrium correlation functions. Profound relations can be obtained from the response theory, such as the fluctuation-dissipation relation~\cite{Callen_PhysRev_1951} and the Green-Kubo formulae~\cite{Green_JChemPhys_1952, Green_JChemPhys_1954, Kubo_JPhysSocJpn_1957a}. From the author's perspective, the response theory is an idea playground for tensor networks to demonstrate their power. In this Letter, we extend the application of tensor networks to examine the response theory with one-dimensional quantum spin systems. For such many-body quantum systems, the underlying Hilbert space can be very huge that the any attempt of numerical calculation with vectors and/or matrices definitely fails. The numerical calculation with tensor networks presented here is very pedagogical, and the result obtained provides a striking illustration of the response theory.

\begin{figure}
\centering
\begin{minipage}[t]{0.8\hsize}
\resizebox{1.0\hsize}{!}{\includegraphics{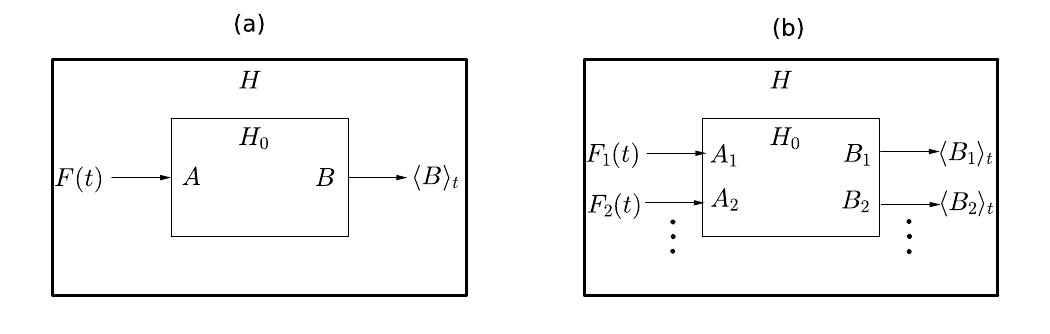}}
\end{minipage}
\caption{Schematic illustration of the responses of properties $\{B_i\}$ induced by time-dependent forces $\{F_i(t)\}$ conjugated with other properties $\{A_i\}$. Panel (a) corresponds to the case of one excitation and one response, while panel (b) the case of multiple excitations and multiple responses.}
\label{fig_response}
\end{figure}

\section{Response Theory}

\par We first give a brief sketch of the basics of the response theory. Let's consider a quantum system that is initially unperturbed with the Hamiltonian $H_0$ and thermally equilibrated with the inverse temperature $\beta\equiv 1/(k_{\rm B}T)$, where $k_{\rm B}$ is the Boltzmann constant. So, the initial density matrix is given by the Gibbs canonical ensemble, $\rho(0)=\rho_{\rm eq}={\rm e}^{-\beta H_0}/Z$, where $Z\equiv{\rm Tr}\left[{\rm e}^{-\beta H_0}\right]$ is the partition function. An excitation is subsequently applied by the time-dependent potential $V(t)=-AF(t)$, where $A$ is the observable and the $F(t)$ the conjugate force. The time evolution of the system's density matrix is then ruled by the von Neumann equation, ${\rm i}\hbar\partial_t\rho(t)=[H(t),\,\rho(t)]$, where $H(t)=H_0+V(t)$ is the total time-dependent Hamiltonian and $[M,N]\equiv MN-NM$ denotes the commutator of two operators. We now switch to the interaction picture, where
\begin{align}
& \rho_{\rm I}(t)\equiv{\rm e}^{+{\rm i}H_0t/\hbar}\rho(t)\,{\rm e}^{-{\rm i}H_0t/\hbar} \text{,} \\
& V_{\rm I}(t)=-A_{\rm I}(t)F(t)\equiv-{\rm e}^{+{\rm i}H_0t/\hbar}A\,{\rm e}^{-{\rm i}H_0t/\hbar}F(t) \text{.}
\end{align}
Then, the von Neumann equation is turned into the similar form
\begin{align}
{\rm i}\hbar\partial_t\rho_{\rm I}(t)=[V_{\rm I}(t),\,\rho_{\rm I}(t)] \text{.}
\end{align}
The average of the interested observable $B$ is calculated as 
\begin{align}
\langle B\rangle_t={\rm Tr}[\rho(t)B]={\rm Tr}[\rho_{\rm I}(t)B_{\rm I}(t)] \text{,}
\end{align}
where $B_{\rm I}(t)\equiv{\rm e}^{+{\rm i}H_0t/\hbar}B\,{\rm e}^{-{\rm i}H_0t/\hbar}$ is the expression in the interaction picture. See Fig.~\ref{fig_response} for the schematic illustration of the responses due to excitations.

\par From the equation of motion for $\rho_{\rm I}(t)$, we get the solution written in Dyson series
\begin{align}
\rho_{\rm I}(t) = & \rho_{\rm I}(0)+\int_0^t\frac{{\rm d}t_1}{{\rm i}\hbar}[V_{\rm I}(t_1),\,\rho_{\rm I}(t_1)] \nonumber \\
= & \rho_{\rm I}(0)+\int_0^t\frac{{\rm d}t_1}{{\rm i}\hbar}[V_{\rm I}(t_1),\,\rho_{\rm I}(0)] \nonumber \\
+ & \int_0^t\frac{{\rm d}t_1}{{\rm i}\hbar}\int_0^{t_1}\frac{{\rm d}t_2}{{\rm i}\hbar}[V_{\rm I}(t_1),[V_{\rm I}(t_2),\rho_{\rm I}(0)]]+\cdots \text{.}
\end{align}
So, we get
\begin{align}
\langle B\rangle_t ={\rm Tr}[\rho_{\rm I}(t)B_{\rm I}(t)] = & {\rm Tr}[\rho_{\rm I}(0)B_{\rm I}(t)]+\frac{1}{{\rm i}\hbar}\int_0^t{\rm d}t_1{\rm Tr}\left\{[V_{\rm I}(t_1),\,\rho_{\rm I}(0)]B_{\rm I}(t)\right\} \nonumber \\
& +\frac{1}{({\rm i}\hbar)^2}\int_0^t{\rm d}t_1\int_0^{t_1}{\rm d}t_2{\rm Tr}\left\{[V_{\rm I}(t_1),\,[V_{\rm I}(t_2),\,\rho_{\rm I}(0)]]B_{\rm I}(t)\right\}+\cdots \text{,} \label{eq_expansion}
\end{align}
which gives terms of zeroth-order, first-order, second-order, and so on. The system is isolated and located in thermodynamic equilibrium at the initial time, $[H_0,\,\rho_{\rm eq}]=0$. As a result, $\rho_{\rm I}(0)=\rho(0)=\rho_{\rm eq}$,
and
\begin{align}
{\rm Tr}[\rho_{\rm I}(0)B_{\rm I}(0)]={\rm Tr}[\rho_{\rm eq}B]=\langle B\rangle_{\rm eq}
\end{align}
is the equilibrium value of the observable $B$. The linear response can be further developed as
\begin{align}
\Delta B_t^{(1)} & =\frac{1}{{\rm i}\hbar}\int_0^t{\rm d}\tau{\rm Tr}\left\{[V_{\rm I}(\tau),\,\rho_{\rm eq}]B_{\rm I}(t)\right\} \nonumber \\
& = \int_0^t{\rm d}\tau\phi_{BA}(t-\tau)F(\tau)  \nonumber \\
& = \int_0^t{\rm d}\tau\phi_{BA}(\tau)F(t-\tau) \text{,} \label{eq_GK_1}
\end{align}
with the linear response function defined by
\begin{align}
\phi_{BA}(\tau) & \equiv\frac{1}{{\rm i}\hbar}{\rm Tr}\left\{[\rho_{\rm eq},\,A]B_{\rm I}(\tau)\right\} \nonumber \\
& = \frac{1}{{\rm i}\hbar}{\rm Tr}\left\{\rho_{\rm eq}[A,\,B_{\rm I}(\tau)]\right\} \text{.}
\end{align}
This function is the equilibrium average of commutator between the observable $A$ at time $0$ and the observable $B$ at time $\tau$. Considering that the equilibrium state is described by the Gibbs canonical ensemble, we have the property
\begin{align}
[{\rm e}^{-\beta H_0},\,A] & ={\rm e}^{-\beta H_0}\int_0^{\beta}{\rm e}^{\lambda H_0}[A,\,H_0]{\rm e}^{-\lambda H_0}{\rm d}\lambda \nonumber \\
& = {\rm i}\hbar{\rm e}^{-\beta H_0}\int_0^{\beta}\dot{A}(-{\rm i}\hbar\lambda){\rm d}\lambda \text{,}
\end{align}
where the first line is obtained by multiplying both side by ${\rm e}^{\beta H_0}$ and then taking derivative with respect to $\beta$. So, the linear response function can be expressed as
\begin{align}
\phi_{BA}(\tau) & =\frac{1}{{\rm i}\hbar}{\rm Tr}\left\{[\rho_{\rm eq},\,A]B_{\rm I}(\tau)\right\} \nonumber \\
& = \int_0^{\beta}{\rm Tr}[\rho_{\rm eq}\dot{A}(-{\rm i}\hbar\lambda)B(\tau)]{\rm d}\lambda \text{.} \label{eq_GK_2}
\end{align}
The Eq.~(\ref{eq_GK_1}) together with Eq.~(\ref{eq_GK_2}) constitute the celebrated Green-Kubo formulae.

\section{Tensor Networks}

\par
\begin{figure}
\centering
\begin{minipage}[t]{0.6\hsize}
\resizebox{1.0\hsize}{!}{\includegraphics{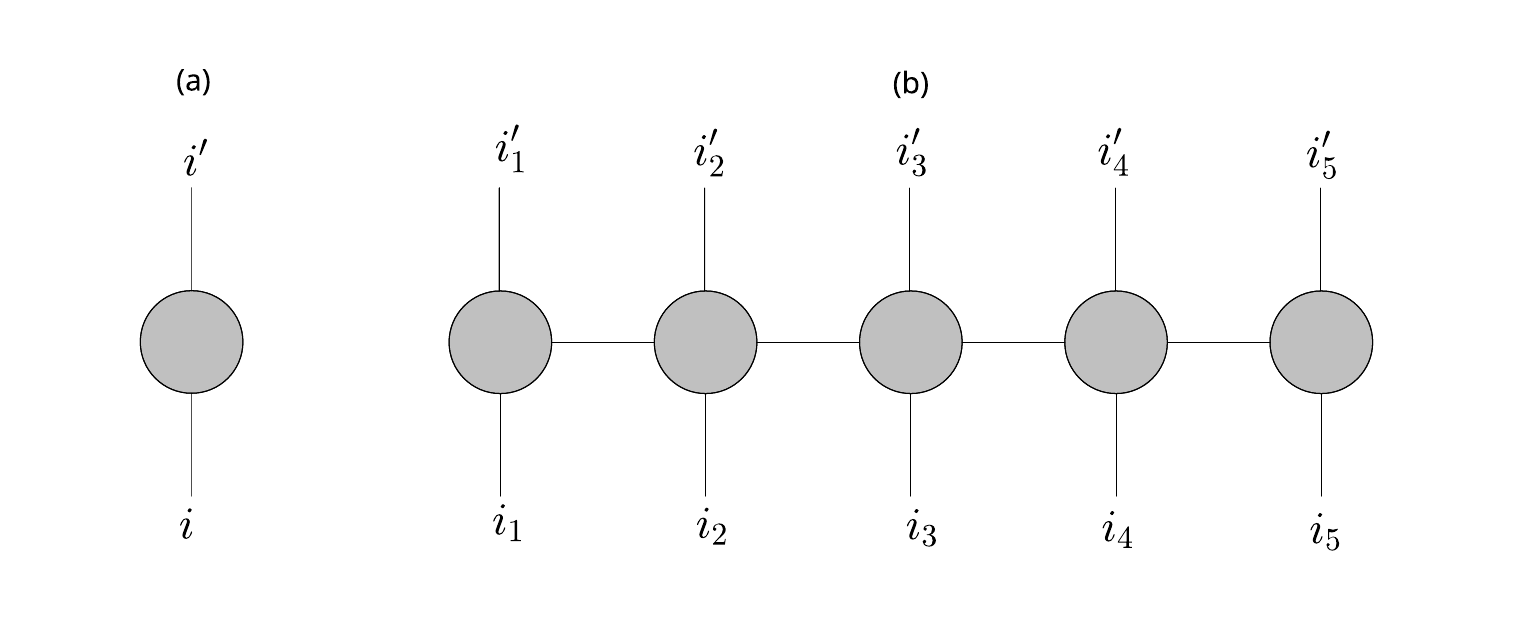}}
\end{minipage}
\caption{Graphical notation of a tensor of order $2$ in panel (a) and a matrix product operator (MPO) in panel (b).}
\label{fig_tensors}
\end{figure}

 The power of tensor networks is exploited to perform numerical calculation. They are the formulation that allows state-of-the-art numerical methods for studying strongly correlated, quantum many-body systems~\cite{Cirac_JPhysAMathGen_2009, Orus_NatRevPhys_2019}. They are especially capable of dealing with one-dimensional quantum system, such as the spin chain discussed here. The outstanding advantage of tensor networks lies in their ability to access large system size. As the name suggests, the building blocks are the tensors, which are array attached with multiple indices. With the terminology in the literature, the number of indices is called order and the number of values each index running over is called dimension. The panel (a) in Fig.~\ref{fig_tensors} shows graphical notation of an order-$2$ tensor that is the equivalence of a matrix. The open legs represent physical indices.

\par For one-dimensional quantum spin chains, the natural extension of order-$2$ tensors are matrix product operators (MPOs), as shown in the panel (b) in Fig.~\ref{fig_tensors}. There are two sets of open legs/indices, one set mapped to the column index of a matrix while the other to the row index. Each index in one set corresponds to the physical degree on each site of the spin chain. A notable difference from an order-$2$ tensor is the shared indices by two neighboring tensors. They are called bond indices, which assume implicit contraction (similar to Einstein summation). This representation of many-body operators significantly reduces the storage needed for the data. The contractions are only performed locally in computation with some tensor network algorithms. After contraction and computation, the resulting large tensors are converted back to some smaller tensors singular value decomposition (SVD).

\par
\begin{figure}
\centering
\begin{minipage}[t]{0.5\hsize}
\resizebox{1.0\hsize}{!}{\includegraphics{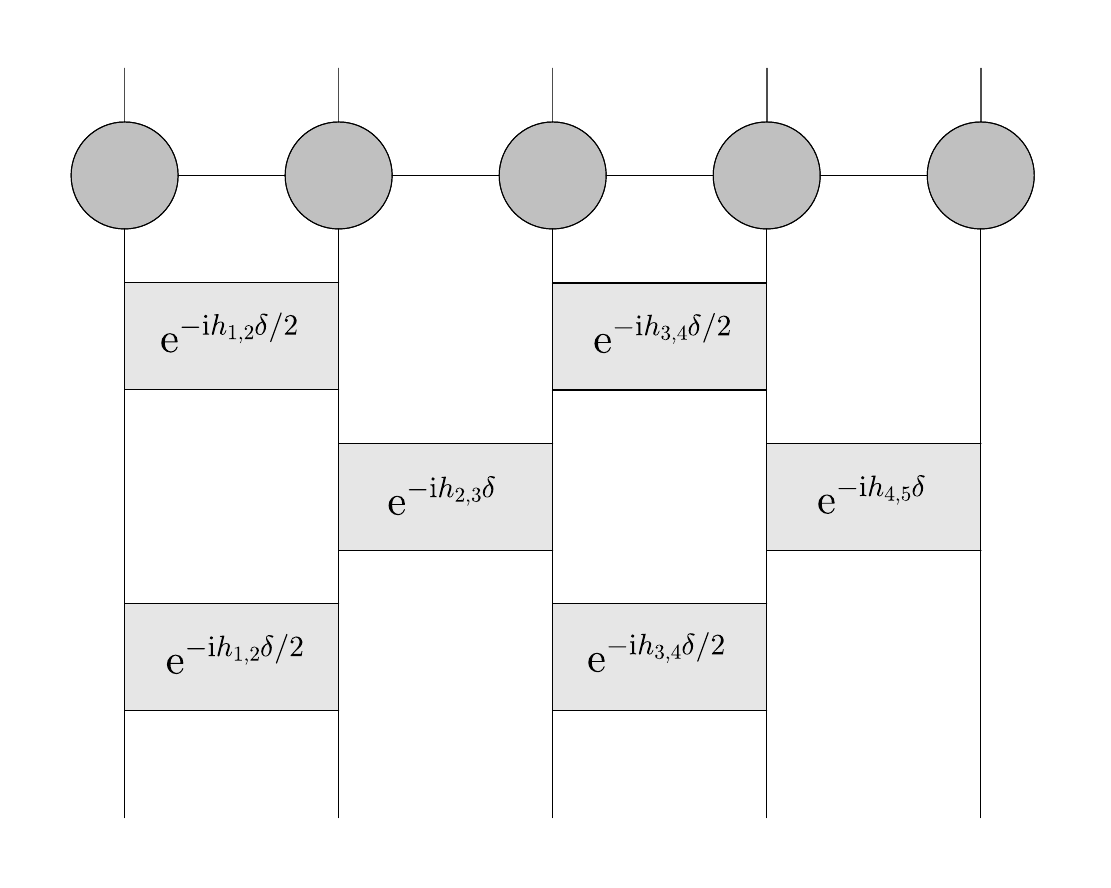}}
\end{minipage}
\caption{Diagrammatic representation of the TEBD algorithm for evolving matrix product operators from one side. Nearest-neighbor interactions between sites are assumed so that the Hamiltonian can be split into odd and even parts, $H=H_{\rm odd}+H_{\rm even}$. Here, $5$ sites are shown for illustrative purpose, and in this case $H_{\rm odd}=h_{1,2}+h_{3,4}$ and $H_{\rm even}=h_{2,3}+h_{4,5}$. Odd and even numbered two-site local evolution operators are alternatively applied.}
\label{fig_TEBD}
\end{figure}

 The tensor network methods needed to be introduced briefly are the real or imaginary time evolution techniques~\cite{Paeckel_AnnPhys_2019}. Time-evolving block decimation (TEBD) is probably the most easy-to-implement simulation method for one-dimensional quantum systems with local interactions~\cite{Vidal_PhysRevLett_2003, Vidal_PhysRevLett_2004}. The underlying idea of this method is to use the Suzuki-Trotter decomposition for a small time-step evolution operator $\exp(-{\rm i}H\delta)$~\cite{Trotter_ProcAmMathSoc_1959, Suzuki_CommunMathPhys_1976}. For each discrete time step, the evolution operator is broken down into a product of operators. In particular, for a nearest-neighbor Hamiltonian $H=\sum_ih_{i,i+1}$, the evolution operator can be written as the second-order decomposition,
\begin{align}
{\rm e}^{-{\rm i}H\delta}\approx {\rm e}^{-{\rm i}H_{\rm o}\delta/2}{\rm e}^{-{\rm i}H_{\rm e}\delta}{\rm e}^{-{\rm i}H_{\rm o}\delta/2}+O(\delta^3) \text{,}
\end{align}
where $H_{\rm o}$ (respectively, $H_{\rm e}$) contains terms $h_{i,i+1}$ with odd (respectively, even) $i$. Consequently, each exponential factor is a product of mutually commuting local terms. See Fig.~\ref{fig_TEBD} for the diagrammatic illustration of the TEBD method applied for the evolution of an MPO. The Time-dependent variational principle (TDVP) represents an alternative method for time evolution~\cite{Haegeman_PhysRevLett_2011, Haegeman_PhysRevB_2016}. Comparatively, it is a little bit more difficult to implement, and no more detailed account here.

\section{One-Dimensional Quantum Spin Chain}

\par For illustrative purpose, we next consider a concrete system -- an spin one-half quantum Ising chain of $L$ sites with nearest-neighbor interactions in the presence of mixed transverse and longitudinal fields. The unperturbed Hamiltonian is given by
\begin{align}
H_0=-J\sum_{\langle i,i+1\rangle}S_i^zS_{i+1}^z-h_x\sum_iS_i^x-h_z\sum_iS_i^z \text{,} \label{eq_H0}
\end{align}
where $S_i^z$ and $S_i^x$ are the spin operators at the $i$-th site defined in terms of Pauli matrices,
\begin{align}
& S_i^x=\frac{\sigma^x}{2}=\frac{1}{2}\begin{pmatrix}
0 & 1 \\
1 & 0
\end{pmatrix} \text{,}
& S_i^z=\frac{\sigma^z}{2}=\frac{1}{2}\begin{pmatrix}
1 & 0 \\
0 & -1
\end{pmatrix} \text{,}
\end{align}
$J$ the coupling constant. $h_x$ (respectively, $h_z$) the magnetic field in $x$- (respectively, $z$-) direction. For this system, it is conveniently to define two observables as the magnetizations in both $z$- and $x$-directions, i.e.,
\begin{align}
S_x=\sum_{i=1}^LS_i^x \text{,}\hspace{1cm} S_z=\sum_{i=1}^LS_i^z \text{.}
\end{align}
The excitation is applied as an additional magnetic field $F(t)$ in $x$-direction so that the time-dependent potential is given by
\begin{align}
V(t)=-S_xF(t) \text{.}
\end{align}
The additional magnetic field is prescribed that it begins to arise from zero at the initial time and finally vanishes at the time $T$, i.e., $F(0)=F(T)=0$. In the following, we numerically calculate the average value at $\langle S_z\rangle$ at time $T$ under the time-dependent Hamiltonian, and compare this to the equilibrium value $\langle S_z\rangle_{\rm eq}$ plus the linear and second-order responses, $\Delta S_z^{(1)}$, $\Delta S_z^{(2)}$. The calculation of the expectation value of the magnetization in $z$-direction at time $T$ is formulated in the Heisenberg picture,
\begin{align}
\langle S_z\rangle_{t=T}={\rm Tr}\left[S_z(T)\rho_{\rm eq}\right] \text{.}
\end{align}
Here, $S_z(T)$ is obtained through the unitary transformation, $S_z(T)=U^{\dagger}S_zU$, where $U$ denotes the unitary evolution operator,
\begin{align}
U\equiv{\cal T}\exp\left[-\frac{\rm i}{\hbar}\int_0^TH(\tau){\rm d}\tau\right] \text{,} \label{eq_U}
\end{align}
expressed in terms of the full time-dependent Hamiltonian $H(t)=H_0+V(t)$ and the time-ordering operator ${\cal T}$.

\begin{figure}
\centering
\begin{minipage}[t]{0.7\hsize}
\resizebox{1.0\hsize}{!}{\includegraphics{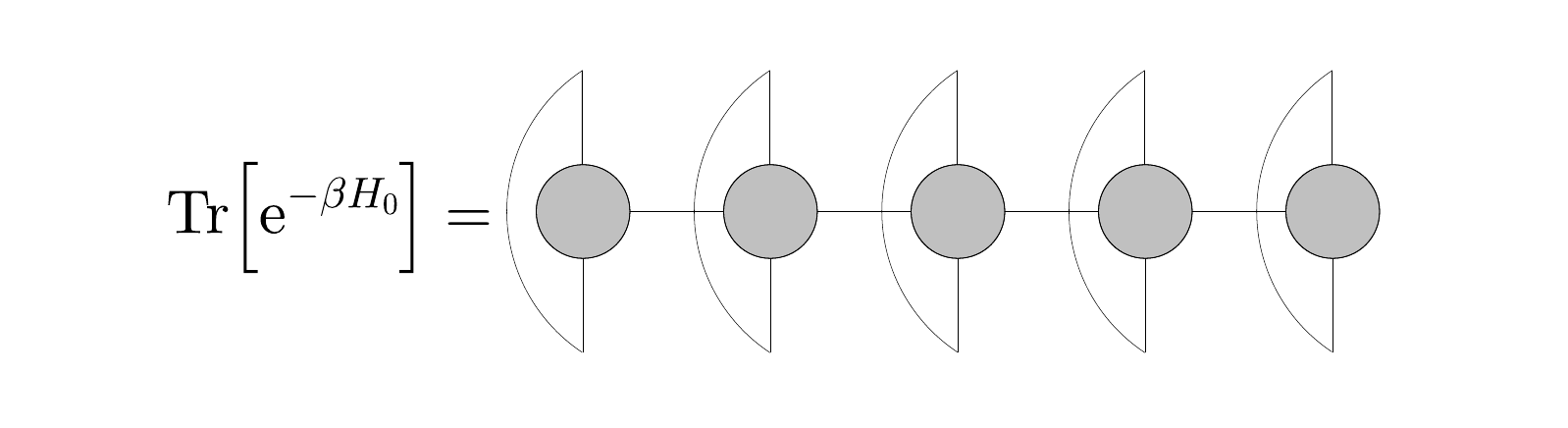}}
\end{minipage}
\caption{Diagrammatic representation of the calculation of the partition function by contracting two sets of physical indices of the MPO representing $\exp(-\beta H_0)$.}
\label{fig_rho}
\end{figure}

\par The common quantity to numerically calculate is the equilibrium density matrix $\rho_{\rm eq}={\rm e}^{-\beta H_0}/Z$. This can be done by imaginary time evolution with TEBD. Firstly, we prepare an initial identity MPO, $\delta_{i_1,i_1'}\delta_{i_2,i_2'}\cdots\delta_{i_L,i_L'}$, where $\{\delta_{i,i'}\}$ are the Kronecker delta defined $1$ if $i=i'$ and $0$ otherwise. Bond indices of dimension $1$ are implicitly assumed to link neighboring Kronecker delta functions. Then, we evolve this identity MPO under $\exp(-\beta H_0)$ with TEBD to obtain the desired matrix product density operator (MPDO) or simply MPO~\cite{Verstraete_PhysRevLett_2004}. The unperturbed Hamiltonian~(\ref{eq_H0}) can be written as the sum of nearest-neighbor terms, $H_0=\sum_ih_{i,i+1}$, with each term constructed as follows
\begin{align}
h_{i,i+1}= & -JS_i^z\otimes S_{i+1}^z \nonumber \\
& -\frac{(1+\delta_{i,1})\cdot(h_xS_i^x\otimes{\sf I}_{i+1})}{2} \nonumber \\
& -\frac{(1+\delta_{i+1,L})\cdot({\sf I}_i\otimes h_xS_{i+1}^x)}{2}  \nonumber \\
& -\frac{(1+\delta_{i,1})\cdot(h_zS_i^z\otimes{\sf I}_{i+1})}{2} \nonumber \\
& -\frac{(1+\delta_{i+1,L})\cdot({\sf I}_i\otimes h_zS_{i+1}^z)}{2} \text{,}
\end{align}
where $\otimes$ stands for the tensor product, and ${\sf I}_i$ the identity operator at the $i$-th site. So, these local terms can be grouped into odd and even categories. The tracing operation in obtaining the partition function is now transformed into tensor contractions of all pairs physical indices, see Fig.~\ref{fig_rho} for diagrammatic illustration. An alternative way to prepare the thermal equilibrium state is by generating a set of typical states representing the Gibbs canonical ensemble. The relevant technique is called minimally entangled typical thermal states (METTS)~\cite{White_PhysRevLett_2009, Stoudenmire_NewJPhys_2010}. The advantage of this technique is that the imaginary time evolution is only up to $\beta/2$ to obtain one typical state. The price to pay is having to sample over many realizations. In practical calculation, the direct evolution of the initial identity MPO by the operator ${\rm e}^{-\beta H_0}$ is adopted.

\begin{figure}
\centering
\begin{minipage}[t]{0.6\hsize}
\resizebox{1.0\hsize}{!}{\includegraphics{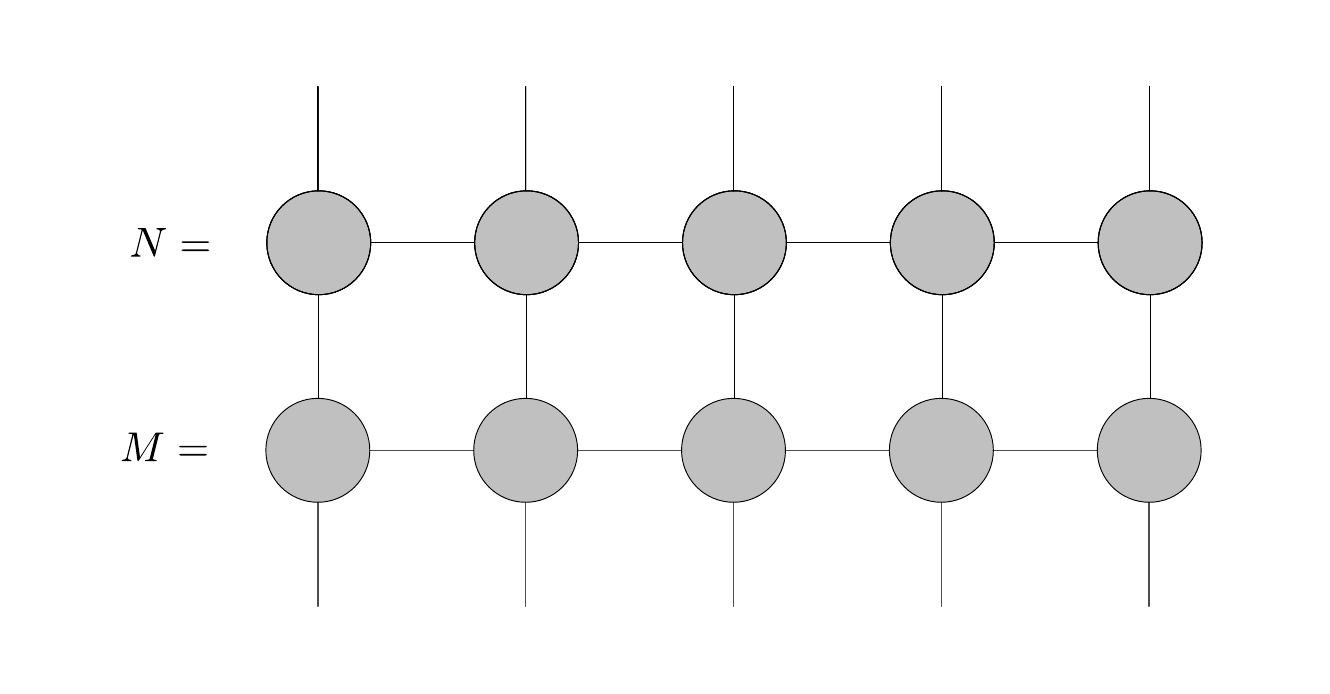}}
\end{minipage}
\caption{Diagrammatic notation of one MPO $M$ multiplied to another MPO $N$. The result is a new MPO $MN$ after contracting the indices between them.}
\label{fig_MPOMPO}
\end{figure}

\par The observable $S_z=\sum_iS_i^z$ can also be represented as an MPO,
\begin{align}
S_z=\begin{pmatrix}
1 & S_1^z
\end{pmatrix}\otimes\cdots\otimes\begin{pmatrix}
1 & S_i^z \\
0 & 1
\end{pmatrix}\otimes\cdots\otimes\begin{pmatrix}
S_L^z \\
1
\end{pmatrix} \text{,}
\end{align}
whose bond indices are all of dimension $2$. The equilibrium value $\langle S_z\rangle_{\rm eq}={\rm Tr}\left[S_z\rho_{\rm eq}\right]$ is numerically calculated by applying the MPO representation of $S_z$ to the MPO representation of $\rho_{\rm eq}$, as illustrated diagrammatically in Fig.~\ref{fig_MPOMPO}, and then taking the trace. For the average value of $S_z$ at time $T$, we should first calculate $S_z(T)=U^{\dagger}S_zU$ with the unitary operator~(\ref{eq_U}). This is done by applying the small time-step evolution operators from both sides. Because the full Hamiltonian is time-dependent, the sequence of these small time-step evolution operators matters. After obtaining $S_z(T)$ in Heisenberg picture, its average can be calculated in the same way as the equilibrium value.

\begin{figure}
\centering
\begin{minipage}[t]{0.6\hsize}
\resizebox{1.0\hsize}{!}{\includegraphics{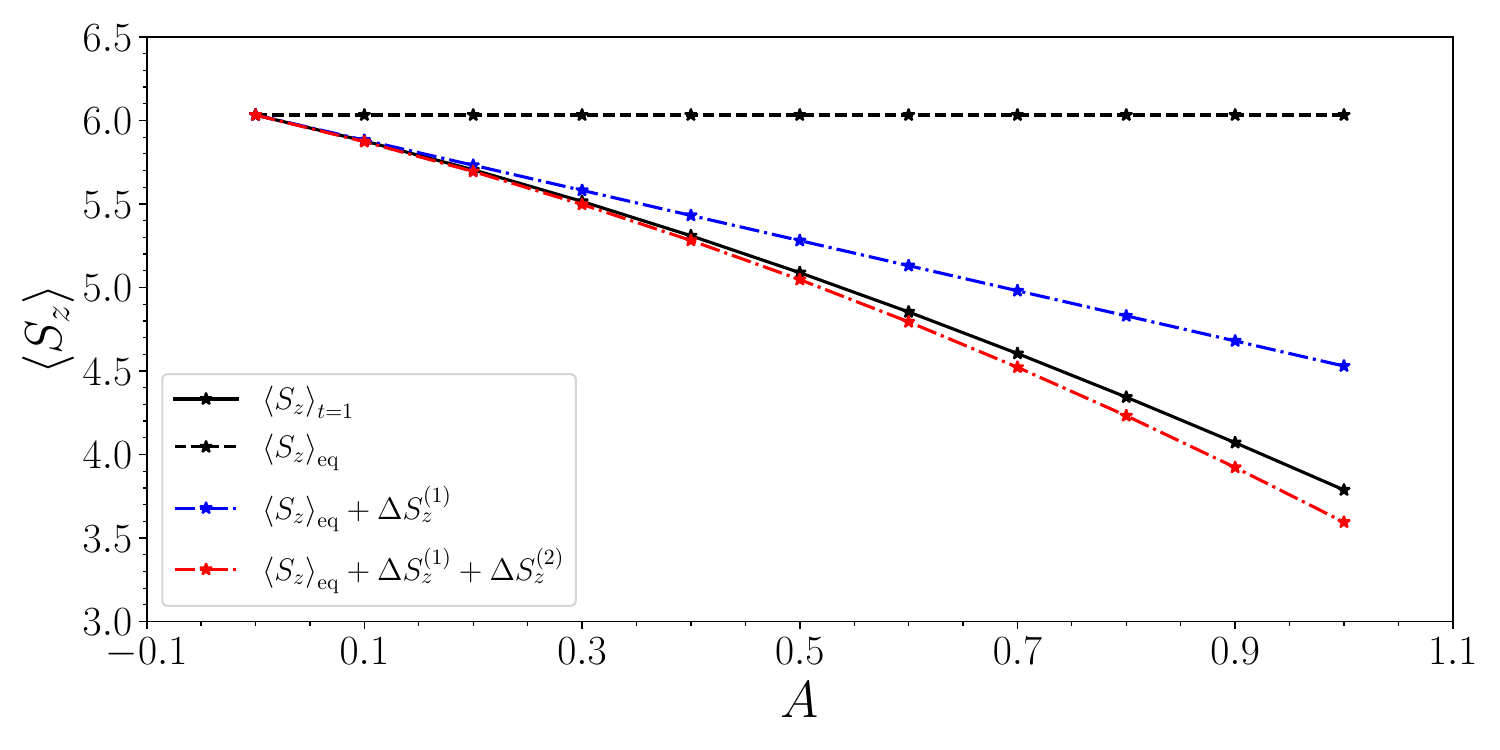}}
\end{minipage}
\caption{The average magnetization in $z$-direction calculated against the amplitude of the excitation. The excitation is taken by the additional field in $x$-direction, $F(t)=A\sin(\pi t)$, where $A$ is the amplitude. The average of $S_z$ is calculated at time $T=1$, at which the additional field vanishes, $F(1)=0$. The system is composed of $20$ sites. The parameter values are $J=h_x=h_z=\hbar=\beta=1$.}
\label{fig_res}
\end{figure}

\par Let's calculate the linear and second-order responses. The former can be calculated from the Green-Kubo formulae~(\ref{eq_GK_1}) and (\ref{eq_GK_2}). It is more direct and convenient to perform calculation from the series expansion~(\ref{eq_expansion}). The time-dependent potential $V_{\rm I}(t)$ and the $S_{z,{\rm I}}(t)$ in the interaction picture are calculated as in the same way stated before
\begin{align}
& V_{\rm I}(t)=-{\rm e}^{+{\rm i}H_0t/\hbar}S_x\,{\rm e}^{-{\rm i}H_0t/\hbar}F(t) \text{,} \\
& S_{z,{\rm I}}(t)={\rm e}^{+{\rm i}H_0t/\hbar}S_z\,{\rm e}^{-{\rm i}H_0t/\hbar} \text{.}
\end{align}
The calculation of responses in Eq.~(\ref{eq_expansion}) involves integral over time, so $V_{\rm I}(t)$ should be calculated at equispaced points of time for numerical integration. The initial density matrix is given by the equilibrium canonical ensemble, $\rho_{\rm I}(0)=\rho_{\rm eq}$. The integrands at difficult time are all calculated with tensor networks.

\par The results are presented in Fig.~\ref{fig_res}, where the equilibrium value $\langle S_z\rangle_{\rm eq}$, the exact average of $S_z$ at time $T$ are plotted against the amplitude of the excitation. The equilibrium value $\langle S_z\rangle_{\rm eq}$ has nothing to do with the excitation, so it is constant. Moreover, the approximate average values of $S_z$ at time $T$ from the equilibrium value plus response corrections are also plotted against the amplitude of the excitation. It is clear that when the amplitude of excitation is small, the linear response is a good correction. However, when the amplitude goes larger, the correction from the second-order response is needed. This expected behavior is very illustrative for the response theory. In the calculation, the system has a total of $20$ sites so that the possible number of states is $1048576$. The underlying Hilbert space is too huge that the exact calculation with normal vectors and/or matrices is impossible. This justifies the need of tensor networks.

\par The computer program for numerical simulation is coded with the ITensor library~\cite{Fishman_SciPostPhysCodeb_2022} in C++. Newcomers are recommended to use the version implemented in Julia language~\cite{Bezanson_SIAMRev_2017}, which is arguably an superior language in nearly every respect in scientific computation.

\section{Conclusion}

\par In the present Letter, the response theory is briefly reviewed and tensor networks are shown to be capable of calculating the responses for one-dimensional quantum spin systems. The Ising chain in mixed transverse and longitudinal fields is chosen as the example. For this system, the excitation is applied to the magnetization in $x$-direction and the response of the $z$-direction magnetization is calculated. The linear and second-order responses are calculated. They are shown to be excellent corrections to the equilibrium magnetization in $z$-direction. This work represents a further application of tensor networks in the field of quantum thermodynamics. Moreover, the result is very illustrative and can serve as an example for educational purpose.

\printbibliography[title={References}]

\end{document}